\documentclass[aps,prb,twocolumn,showpacs,superscriptaddress,floatfix]{revtex4}
\usepackage{epsfig,amsmath,amssymb}
\def\etal#1{#1}

\newcommand{\be}{\begin{equation}}
\newcommand{\ee}{\end{equation}}

\newcommand{\bit}{\begin{itemize}}
\newcommand{\eit}{\end{itemize}}
\newcommand{\bea}{\begin{eqnarray}}
\newcommand{\eea}{\end{eqnarray}}
\newcommand{\stateX}{\(\sqrt{3}\)\(\times\)\(\sqrt{3}\)}
\newcommand{\stateO}{\(q\)\(=\)\(0\)}
\newcommand{\neel}{N\'{e}el}
\newcommand{\Neel}{N\'{e}el }
\newcommand{\Kagome}{kagom\'e}
\newcommand{\kagome}{{\Kagome} }
\newcommand{\ebond}{e_0}
\newcommand{\Sz}{{\hat S}^z}
\begin{document}

\title{Absence of magnetic order for the spin-half
Heisenberg antiferromagnet on the star lattice}

\author{J. Richter}
\affiliation{Institut f\"ur Theoretische Physik, Universit\"at Magdeburg,
      P.O. Box 4120, D-39016 Magdeburg, Germany}
\author{J. Schulenburg}
\affiliation{Universit\"{a}tsrechenzentrum,
             Universit\"{a}t Magdeburg,
             P.O. Box 4120, D-39016 Magdeburg, Germany}
\author{A. Honecker}
\affiliation{Institut f\"ur Theoretische Physik, TU Braunschweig,
      D-38106 Braunschweig, Germany}
\author{D. Schmalfu{\ss}}
\affiliation{Institut f\"ur Theoretische Physik, Universit\"at Magdeburg,
      P.O. Box 4120, D-39016 Magdeburg, Germany}

\date{June 2, 2004; revised September 13, 2004}

\begin{abstract}
We study the ground-state properties of the spin-half Heisenberg
antiferromagnet on the two-dimensional star lattice by spin-wave theory,
exact diagonalization and a variational mean-field approach.
We find evidence that the star
lattice is (besides the \kagome lattice) a second candidate among the 11
uniform Archimedean lattices where quantum fluctuations in
combination with frustration lead to a quantum paramagnetic ground state.
Although the classical ground state of the Heisenberg antiferromagnet
on the star lattice exhibits a huge non-trivial degeneracy like on the
\kagome lattice, its quantum ground state is most likely dimerized
with a gap to all excitations. Finally, we find several candidates
for plateaux in the magnetization curve as well as a macroscopic magnetization
jump to saturation due to independent localized magnon states.
\end{abstract}


\maketitle

\section{Introduction} 
The spin-half two-dimensional (2D) Heisenberg antiferromagnet (HAFM) has 
attracted much attention in recent times.  
In particular, the recent progress in synthesizing novel 
quasi-2D magnetic materials exhibiting exciting quantum effects has
stimulated much theoretical work.  
We mention for example the spin-gap behavior in CaV$_4$O$_9$ \cite{taniguchi95}
and in SrCu$_2$(BO$_3$)$_2$,\cite{kageyama99} 
the spin fractionalization in Cs$_2$CuCl$_4$ \cite{coldea}
or the plateau structure in the magnetization
process of frustrated quasi-2D magnetic materials like
SrCu$_2$(BO$_3$)$_2$ \cite{kageyama99}
or Cs$_2$CuBr$_4$.\cite{TOKMIG02}

While the ground state (GS) of the one-dimensional quantum HAFM does not
possess \Neel long-range order, for the spin-half HAFM on 2D lattices the
competition between quantum fluctuations and interactions seems to be well
balanced and magnetically ordered and disordered GS phases may appear.
A fine tuning of this competition 
may lead to zero temperature  transitions between
semi-classical and quantum phases.
The prototypes of 2D arrangements of spins 
are the 11 uniform Archimedean lattices (tilings), see, e.g.\
Refs.\ \onlinecite{gruenbaum,phd,wir04}.
They present an ideal possibility
for a systematic study of the interplay of   
lattice geometry and magnetic interactions in 2D quantum spin systems. 

The HAFM on the widely known  square, honeycomb, triangular and \kagome
lattices has been  studied in numerous papers
over the last decade.
While for the square, honeycomb and   triangular lattices the existence of
semi-classical magnetic order seems to be well-established (see e.g.\
Refs.\ \onlinecite{wir04,lhuillier03,lhuillier01})
the spin-half HAFM on the \kagome
lattice is a candidate for a magnetic system with a quantum paramagnetic GS 
(see the reviews \onlinecite{wir04,lhuillier03,lhuillier01,moessner01} and 
references therein). Other less
known Archimedean lattices like the maple-leaf lattice,\cite{schmalfuss02}
the square-hexagonal-dodecagonal 
lattice \cite{tomczak99,tomczak01} and the trellis lattice
\cite{normand97,miyahara98} 
exhibit most likely semi-classical magnetic GS order.

\begin{figure}[t!]
\centerline{\epsfig{file=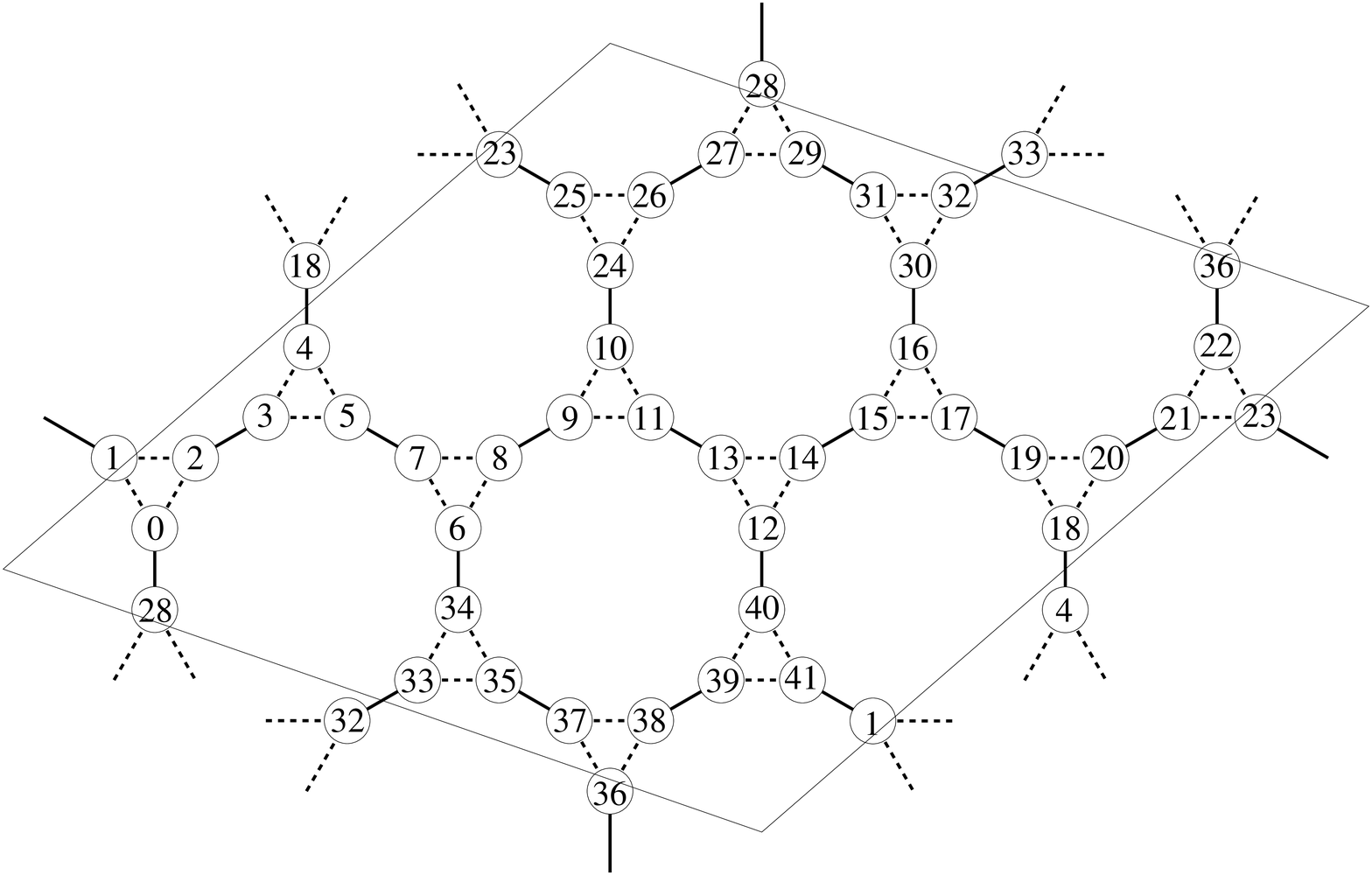,width=\columnwidth}}
\caption[]{
The star lattice with $N=42$ sites.} 
\label{fig1}
\end{figure}

In this paper we present another candidate for a 
quantum paramagnetic GS among the Archimedean 
lattices, namely the so-called star lattice, featured by low coordination
number $z=3$ and strong frustration due to a triangular arrangement of bonds
(see Fig.~\ref{fig1}).
%

\section{Model}
The geometric unit cell of the star lattice contains six 
sites and  the underlying Bravais lattice is a triangular one
(see Figs.\ \ref{fig1} and \ref{fig1a}).
For this lattice we consider the spin-half HAFM in a  magnetic field $h$
\begin{equation}
\label{Ham1}
 \hat{H} = J \sum_{\langle ij \rangle}{\bf S}_i \cdot {\bf S}_j
- h \hat{S}^z,
\end{equation}
where the sum runs over pairs of neighboring sites $\langle ij \rangle$ and 
$\hat{S}^z = \sum_i\hat{S}^z_i$.  
The star lattice carries topologically inequivalent nearest-neighbor (NN) 
bonds $J_D$ (dimer bonds, solid lines in Fig.~\ref{fig1}) and $J_T$
(triangular bonds, dashed lines in Fig.~\ref{fig1}, see also Fig.~\ref{fig1a}).
For the uniform lattice these bonds are of equal strength $J_D=J_T=J$.

\section{Semi-classical ground state}
\begin{figure}[t!]
\centerline{\epsfig{file=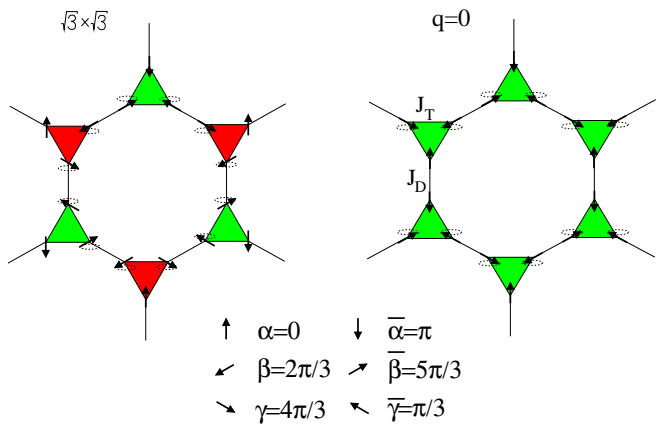,width=1.\columnwidth}}
\caption[]{(Color online)
Two variants of the GS of the classical HAFM on the star lattice:
the \stateX$ $ state (left) and the \stateO$ $ state (right).
The dotted ellipses show further degrees of freedom of the highly
degenerate classical GS.
Different shades of the triangles symbolize different chiralities.
}
\label{fig1a}
\end{figure}

In the classical GS for $h=0$ the two non-equivalent NN bonds of the
star lattice
carry different NN spin correlations: The angle between neighboring spins 
on dimer bonds $J_D$ is $\pi$, 
whereas the angle on triangular bonds $J_T$ is $2 \pi/3$. 
Its energy per bond is $\ebond^{\rm class} =-1/6$.
The classical GS for the star lattice has a great similarity to that of the
\kagome lattice. It also exhibits 
a non-trivial infinite degeneracy. Moreover, there are also the 
two variants of the classical GS, namely the
so-called \stateX $\mbox{}$ and
$q=0$ states (see Fig.\ \ref{fig1a}),
often used for discussing possible order in the \kagome
lattice. Hence these two particular planar states can also be considered 
as variants of possible GS ordering for the star lattice.
In the following we discuss the influence of quantum fluctuations on the GS
properties.


First, we perform a linear spin-wave theory (LSWT)
starting from the planar classical GSs.
The LSWT for the star lattice is more ambitious than for the
kagom\'{e} lattice, since we have to consider six types of magnons.
As in the kagom\'{e} case \cite{harris92,ChaHoShe,asakawa94} the
spin-wave spectra are equivalent for all coplanar 
configurations satisfying the classical GS constraint.
We obtain six spin-wave branches.
Two dispersionless modes are found, namely 
$\omega_{1{\bf{q}}}=0,\omega_{2{\bf{q}}}=Js\sqrt{3}$. 
Thus also a flat zero-mode appears as it is observed for the \kagome case.
In addition there are two acoustical and two optical branches.
There is no {\it order-by-disorder} selection among the coplanar classical
GSs due to the equivalence of the spin-wave branches obtained from
LSWT, exactly like for the \kagome lattice.\cite{ChaHoShe,moessner01}
The GS energy per bond for $s=1/2$ in LSWT is $\ebond/J = -0.296759$.
Due to the flat zero mode the integral for the sublattice 
magnetization diverges \cite{asakawa94} which might be understood
as another hint for the absence of the classical order. 
Although on the semi-classical LSWT level both the \kagome and the star lattice
exhibit almost identical properties, the situation might be changed taking
into account the quantum fluctuations more properly. 

\section{Lanczos exact diagonalization}
\label{secLanc}
We consider now the extreme quantum limit $s=1/2$ by direct numerical 
calculation of the GS and the low-lying excitations at $h=0$ for finite
lattices of $N=18, 24, 30, 36, 42$ sites. For each size $N$ we have chosen
only lattices having good geometric properties using the criteria 
given in Ref.\ \onlinecite{betts98}.
The largest lattice ($N=42$) is shown in Fig.~\ref{fig1} and required
a Lanczos diagonalization in dimension 801~258~898 for the computation
of GS properties.
The GS for all these lattices is a singlet and has an energy per bond
$\ebond =  -0.312479$ ($N=18$); 
$-0.311342$ ($24$); 
$-0.310808$ ($30$);  
$-0.310348$ ($36$A); 
$-0.310657$ ($36$B);
$-0.309918$ ($42$). 
The first excitation is a triplet and has a gap 
$\Delta = 0.578710$ ($N=18$); 
$0.531822$ ($24$); 
$0.498564$ ($30$);  
$0.480343$ ($36$A);
$0.483112$ ($36$B)
(no result available for $N=42$).

We present in Table \ref{cor42} the spin-spin correlation 
for the largest finite lattice considered.
Note that  the two non-equivalent NN correlation functions 
differ drastically, we have 
$\langle \Sz_0 \Sz_{28}\rangle  \sim 3.5 
\langle \Sz_0 \Sz_{1}\rangle$ 
indicating a tendency to form local singlets on the dimer bonds.

\begin{table}[b]
\caption[star $N=42$]{\label{cor42} All non-equivalent 
GS spin-spin cor\-rela\-tions $\langle \Sz_0 \Sz_{i}\rangle= \frac{1}{3}
\langle {\bf S}_0{\bf S}_{i} \rangle$  for the  
HAFM on the star lattice with  $N=42$ sites.
In addition to the site index $i$ we give the separation $R=|{\bf R}_0 -
{\bf R}_i|$ between sites $0$ and $i$ in units of NN separation. 
 }
\begin{center}
\begin{tabular}{r|r|r|r}
\hline
$ i  \; \;(R) \quad$                   &  1  $\; \;(1) \quad$  &  3 $\; \;(1.932) \;$ & \;4 $ \; \;(2.909) \;$  \\ 
$\langle \Sz_0 \Sz_{i}\rangle   \; $   & $-$0.05643  $\; $      &  0.03559  $\; $     & $-$0.01058 $\;$         \\ \hline 
$ i  \; \;(R) \quad$                   & 6 $\; \;(3.732) \; $  &  7 $\; \;(3.346) \;$ & 9 $\;  \;(5.278) \;$   \\ 
$\langle \Sz_0 \Sz_{i}\rangle  \; $    & $-$0.00451   $\; $     &  0.01066 $\;$       & 0.00439 $\; $         \\ \hline
$ i  \; \; (R) \quad $                 & 10 $\; \;(4.625) \;$  & 26 $\; \;(2.732) \;$ & 28 $\;    \;(1) \;$  \\ 
$\langle \Sz_0 \Sz_{i}\rangle  \; $    & $-$0.01173 $\;$        & $-$0.03875 $\; $      & $-$0.19707  $\; $         \\ \hline
\end{tabular}
\end{center}
\end{table}
%

\begin{figure}[t!]
\centerline{\epsfig{file=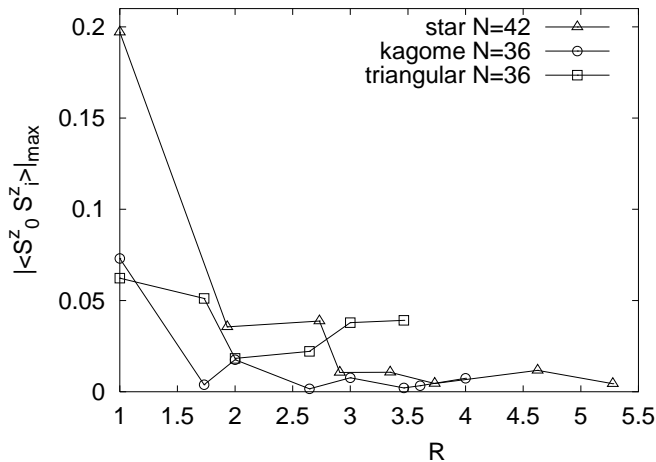,angle=270,width=\columnwidth}}
\caption[]{
Maximal spin-spin correlation
$|\langle \hat{S}^z_0 \hat{S}_i^z\rangle|_{\max}$ 
versus separation $R=|{\bf R}_0 - {\bf R}_i|$ for the star lattice ($N=42$),
the \kagome ($N=36$) and the triangular lattice ($N=36$)
(the lines are guides for the eyes).
The data for the kagom\'e lattice coincide with those from
Ref.~\onlinecite{leung93}. 
}
\label{fig2}
\end{figure}

Let us compare the spin correlations with 
those for the HAFM on triangular and kagom\'e lattices.
We consider the strongest correlations as a measure for
magnetic order and present in Fig.~\ref{fig2}
the maximal absolute correlations
$|\langle \hat{S}^z_0 \hat{S}^z_i \rangle|_{\max}$
for a certain separation $R=|{\bf R}_0 - {\bf R}_i|$ versus $R$. 
As expected we have very rapidly decaying correlations for the disordered 
kagom\'e case, whereas the correlations for the N\'{e}el ordered
triangular lattice are much stronger for larger distances and 
show a kind of saturation for larger $R$.
Although the correlations for the star lattice are slightly larger than 
those of the
\kagome lattice they are significantly weaker for separations $R \ge 3$ than 
those of the triangular lattice.
The large NN correlation for the star lattice 
corresponds to a NN dimer bond.

\begin{figure}[t!]
\centerline{\epsfig{file=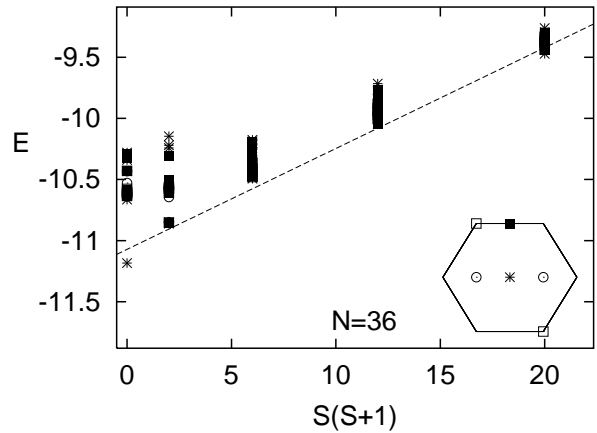,angle=0,width=\columnwidth}}
\caption[Pisa]{ 
Low-energy spectrum for the HAFM on the star lattice ($N=36$B)
(the inset shows the {\bf k} points in the Brillouin zone).
}
\label{fig3}
\end{figure}

Next we consider the 
low-lying spectrum of the star lattice (see Fig.~\ref{fig3}), 
following the lines of the discussion of the spectrum for the triangular
\cite{bernu94} and the \kagome lattice.\cite{lecheminant97,waldtmann98}
It is obvious
that the lowest states $E_{\min}(S)$ are not well described by 
the effective low-energy Hamiltonian $H_{\rm eff} \sim E_0 + {\bf S}^2/2N\chi_0$
of a semi-classically ordered system:
(i) The $E_{\min}$ versus $S(S+1)$ curve 
deviates significantly from a straight line (cf.\ the dashed line 
in Fig.~\ref{fig3}). 
(ii) We do not see well separated 
lowest states in the different sectors of $S$ (so-called 
quasi degenerate joint states \cite{bernu94}) 
which could collapse onto
a \neel-like state in the thermodynamic limit.
(iii) The symmetries of the lowest states in
each sector of $S$ cannot be attributed to the classical \stateX$ $ 
or \stateO$ $ GSs in general. These features are similar to the
\kagome lattice.\cite{lecheminant97,waldtmann98}
However, there is one striking difference. In contrast to the \kagome
lattice we do not have 
non-magnetic singlets filling the singlet-triplet gap (spin gap)
commonly interpreted as a remnant of the
huge classical GS degeneracy. Rather we have a particularly large
singlet-singlet gap. 
This basic difference to the \kagome lattice can be attributed to the 
special property of the quantum GS of the star lattice to have strongly
enhanced antiferromagnetic correlations on the $J_D$ bonds.
As a consequence the quantum GS of the star
lattice has an exceptionally low GS energy $e_0$ (see Table \ref{tab2})
and is well separated from all excitations.

\begin{table}[b]
\caption[Vergleich]{ \label{tab2}
Results of the finite-size extrapolation of the GS energy per bond
$\ebond$ and the order parameter $m^+$ (eq.~(\protect{\ref{mdef}}))
of the spin-half HAFM for the star ($N=18,24,30,36,42$), the \kagome
($N=12,18,24,30,36$) and the triangular ($N=24,30,36$) lattices.
To see the effect of quantum fluctuations we scale $m^+$ by its
corresponding classical value $m^+_{\rm class}$. 
}
\begin{center}
\begin{tabular}{l|c|c|c}
 lattice               & triangular & \kagome   & star   \\ \hline
 $\ebond$              & $-0.1842$  & $-0.2172$ & $-0.3091$ \\
 $m^+/m^+_{\rm class}$ & $0.386$    & $0.000$   & $0.122$ 
\end{tabular}
\end{center}
\end{table}

For finite systems the order parameter is based on the spin-spin correlation
functions. For systems with well-defined semi-classical 
long-range order usually the square of the staggered magnetization is used.
However, this definition of the order parameter
is problematic in the present situation: due to the
huge non-trivial degeneracy of the classical GS it remains unclear 
which type of ordering might be favored in the quantum system.
Therefore we use a definition of an order parameter
\begin{equation} \label{mdef}
  m^+ =\Big(\frac{1}{N^2} \sum_{i,j}
  {}|\langle{\bf S}_{i}{\bf S}_{j}\rangle|\Big)^{1/2}
\end{equation}
which is independent of any assumption on classical order.\cite{wir04}
For bipartite systems like the square lattice 
this definition is identical to the
staggered magnetization $\bar m$ and for 
the HAFM on the triangular lattice
$(m^+)^2$ is by $1/3$ larger than the usual definition.\cite{wir04}
For  the two planar classical \stateX $\mbox{ }$ and  \stateO 
$\mbox{ }$ GSs we get $m^+_{{\rm class},\sqrt{3}\times\sqrt{3}}=
m^+_{{\rm class},q=0}=\frac{1}{2}\sqrt{2/3} \approx 0.40825$
(note that for the \kagome lattice one obtains the same value).
For the quantum model one finds 
$(m^+)^2=0.149113$ ($N=18$); $0.114822$ ($24$);
$0.094831$ ($30$); $0.082299$ ($36$A); $0.079351$ ($36$B);
$0.073251$ ($42$).
For comparison we quote the values for the $N=36$ \kagome lattice:
$(m^+)^2=0.059128$, and the $N=36$ triangular lattice: $(m^+)^2=0.124802$. 

We have performed finite-size extrapolations 
based on the standard formulas for 2D spin-half Heisenberg antiferromagnets
(see, e.g.\ Refs.\ \onlinecite{wir04,neuberger89,hasenfratz93}), namely 
$e_0(L)=  A_0 + \frac{A_3 }{L^{3}} + {\cal O}(L^{-4}) 
$ 
for the GS energy per bond,
$m^+(L) = B_0 + \frac{B_1 }{L} +  {\cal O}(L^{-2})
$
for the order parameter, and
$
\Delta(L) = G_0 +  \frac{G_2 }{L^2}  + {\cal O}(L^{-3}) 
$
for the spin gap, where
$A_0=e_0(\infty)$,   
$B_0=m^+(\infty)$,
$G_0=\Delta(\infty)$ and 
$L=N^{1/2}$.
We present the results of the extrapolation for $\ebond$ and $m^+$
in comparison with the triangular and the \kagome lattice in Table \ref{tab2}.
The HAFM on the star lattice has lowest GS energy $e_0$;
the extrapolated order parameter is finite but very small.

\begin{figure}[t!]
\centerline{\epsfig{file=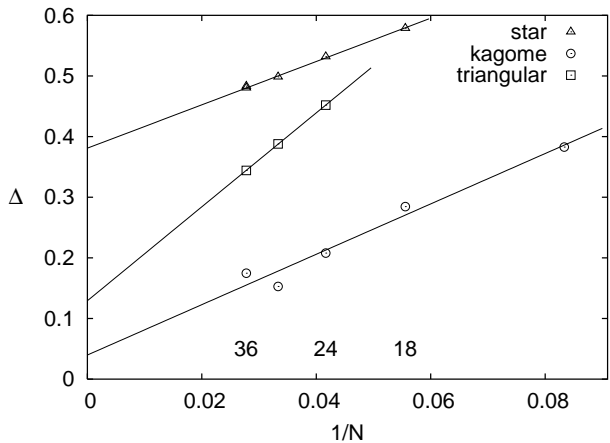,width=\columnwidth}}
\caption[]{
Finite-size behavior of the spin gap $\Delta$, i.e.\ the gap to
$s=1$ excitations versus inverse
system size $1/N$. Results are shown for the star lattice (triangles)
in comparison with those for the \kagome lattice (circles,
compare Ref.\ \onlinecite{waldtmann98}) and
the triangular lattice (squares).
}
\label{figGap}
\end{figure}

Fig.\ \ref{figGap} shows the finite-size behavior of the spin gap of
the star lattice in comparison to the \kagome and triangular lattices.
We extrapolate a quite big spin gap $\Delta = 0.380$ for the star lattice.
For the triangular lattice, the data with $N \le 36$ seem to suggest
a finite spin gap which is, however, spurious.
This illustrates that the extrapolation of the spin gap may be most affected by
systematic errors.\cite{wir04}
Nevertheless, the spin gap of the star lattice
in Fig.\ \ref{figGap} exhibits only comparably small finite-size effects.
Hence, the estimation of a non-zero spin gap for the
star lattice seems to be reliable.  
The spin gap extrapolated for the \kagome lattice \cite{lhuillier03} 
is more than six times smaller, but note that the
existence of a gap for the \kagome lattice at all is not fully clear,
as is also evident from Fig.\ \ref{figGap}.

\section{Variational mean field approach}
We discuss briefly a variational approach which
was successfully applied to describe a quantum phase transition  
between \Neel phases and a dimer phase.\cite{wir04,gwr95,krsfb00}
Let us consider a model with different NN bonds
$J_T$ and $J_D$. The GS shall be approximately described 
by a variational wave function 
\begin{equation}
\label{var}
|\Psi\rangle=
\prod_{\alpha } 
\frac{|\phi^i_+(\theta_\alpha)\rangle|\phi^j_-(\theta_\alpha)\rangle
-t |\phi^i_-(\theta_\alpha)\rangle|\phi^j_+(\theta_\alpha)\rangle}
{\sqrt{1+t^2}}
\end{equation}
where $\alpha$ represents a pair of sites $i,j$ corresponding to a $J_D$ bond. 
Thus the product in (\ref{var}) is effectively taken over all $J_D$ bonds of the
star lattice. In (\ref{var}) the
vectors $|\phi^i_{\pm}(\theta_\alpha)\rangle$ are spin up (down)
states at site $i$ with a quantization axis corresponding to 
the classical planar GS considered, i.e.\ 
$[\sin(\theta_\alpha)\hat{S}_i^x +
\cos(\theta_\alpha)\hat{S}_i^z ]  
|\phi^i_{\pm}(\theta_\alpha)\rangle = \pm\frac{1}{2}  
|\phi^i_{\pm}(\theta_\alpha)\rangle$, where
the angles $\theta_\alpha$ correspond to the respective 
classical \stateX$ $
or \stateO$ $ state.
$|\Psi\rangle$ depends on the variational
parameter $t$ and interpolates between a rotationally invariant 
dimer product state for $t=1$ and an uncorrelated planar \stateX$ $
or \stateO$ $ state for $t=0$.
Optimizing $\langle\Psi|\hat{H}|\Psi\rangle$ with respect to $t$ we get 
$
 E^{\rm var}_0/\mbox{bond} = -(J_D^2+J_D J_T+J_T^2)/12J_T
$.
For the sublattice magnetization we obtain 
\be
\label{eq11}
M= \langle\Psi|\cos(\theta_\alpha) \hat{S}_i^z
+ \sin(\theta_\alpha) \hat{S}_i^x |\Psi\rangle
= \frac{\sqrt{J_T^2-J_D^2}}{2 J_T}
\ee
for $J_D \le J_T$. $M$ vanishes with a mean-field critical exponent at 
the symmetric point $J_D=J_T$.
Since such an approach tends to overestimate the region of the semi-classically
ordered state,\cite{gwr95,krsfb00} we may interpret the above result as
a further indication of a dimerized GS.

\section{Magnetization process}
\begin{figure}[t!]
\centerline{\epsfig{file=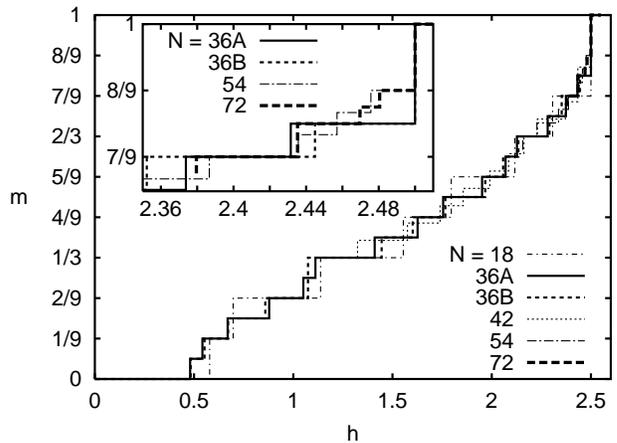,angle=270,width=1.0\columnwidth}}
\caption[]{
Magnetization curves for star lattices with $N=18$, $36$A and $36$B (complete)
and $N=42$, $54$ and $72$ (partial).
Inset: High-field part of the magnetization curves for
$N=36$A, $36$B, $54$ and $72$ sites.
}
\label{fig4}
\end{figure}

Finally, Fig.~\ref{fig4} shows magnetization curves of several finite
star lattices, where the magnetization $m$ is defined as
$m= 2 \langle {\hat S}^z \rangle/N$.
Due to computational limitations, only the high-field region can
be studied for $N > 36$. For example for $N=42$, reliable data are available
only for $m \ge 1/3$ (and of course $m=0$, see section \ref{secLanc}).
Furthermore, the lowest parts of the curves for $N=54$ ($17/27 \le m < 19/27$)
and $N=72$ ($3/4 \le m < 5/6$) are based on assumptions concerning the symmetry
of the GS.

First, one observes a pronounced zero-field plateau in Fig.~\ref{fig4}
corresponding to the spin gap discussed in section \ref{secLanc}.
Candidates for further plateaux emerge e.g.\ at $m=1/3$, $7/9$ and $8/9$.
The finite-size effects at the boundaries of these candidate
plateaux are comparably weak for $m=7/9$ (see inset of Fig.\ \ref{fig4})
such that the numerical evidence in favor of a plateau with $m \ne 0$
is strongest in this case.

The presence of a plateau at $m=1/3$ is also
plausible since the star lattice consists of triangles. More precisely,
Ising-like anisotropies can be argued to give rise to a plateau
at $m=1/3$ due to up-up-down configurations on the
triangles. The number ${\cal N}_{\rm conf.}$ of 
such Ising configurations can be determined by explicit enumeration,
yielding e.g.\  ${\cal N}_{\rm conf.} =$ 123 528, 3 508 392, and
531 606 684 for the $N=$42, 54, and 72 lattices, respectively. This
number grows asymptotically approximately as
${\cal N}_{\rm conf.} \propto (1.322\ldots)^N$. The number of
Ising configurations is much bigger than the corresponding
number on the \kagome lattice at $m=1/3$ (see Ref.\ \onlinecite{kagplateau}
and references therein), indicating that the tendency towards a
disordered GS at $m=1/3$ may be stronger on the star lattice than
on the \kagome lattice.

Just below saturation, we see a jump in the magnetization curve
(see inset of Fig.~\ref{fig4}). Indeed, the presence of this jump
follows from a general construction of independent localized magnons
for strongly frustrated lattices,\cite{jump,wir04}
which in the case of the star lattice live on the dodecagons.
The expected height $\delta m=1/9$ of the jump for sufficiently large $N$
is confirmed for $N=54$ and $72$ (see inset of Fig.~\ref{fig4}).
Note that the existence of localized magnon states also leads to a finite
residual $T=0$ entropy at the saturation field $h=5 J/2$\cite{wir04,entropy}
and may favor a tendency towards a spin-Peierls deformation.\cite{sp04}
On general grounds one expects a plateau just below the jump, i.e.\
at the candidate value $m=8/9$ mentioned before,\cite{wir04} although
the available data do not allow unambiguous confirmation of this plateau.

\section{Discussion and Conclusion}
Similar as for the \kagome lattice the results reported in this paper
yield indications for a quantum paramagnetic GS for the star lattice, too.
Whereas this statement is well-known for the \kagome lattice,   
the star lattice represents a new example for a quantum HAFM  on a
uniform 2D lattice without semi-classical GS ordering. 
However, we emphasize that despite the fact that on the classical and
semi-classical level (LSWT) we have very similar
physics as for the \kagome lattice (i.e.\ one zero mode, the
classical GS degeneracy is not lifted) 
the quantum paramagnetic GS for the star
lattice is of different nature than that for the \kagome lattice. 
The quantum GS for the star lattice is characterized by extremely strong NN
correlation on the dimer bonds (more than 60\% larger than the NN correlation 
of the honeycomb lattice having the same coordination number $z=3$)
and a weak NN correlation on the triangular bonds (only about 30\% of the
NN dimer correlation and
significantly less than the triangular NN correlation 
of the \kagome and the triangular lattices).
The singlet-triplet spin gap is particularly large (about six times larger
than that for the \kagome lattice). 
Although the classical GS exhibits a huge non-trivial degeneracy remarkably 
one does not find low-lying singlets within this large spin gap,  
rather the first singlet excitation is well above the lowest triplet state.
The low-lying spectrum as a whole resembles the spectrum of weakly coupled 
dimers.\cite{phd}
All these features support the conclusion that the quantum GS of the HAFM on
the star lattice is dominated by local singlet pairing. This dimerized GS
represents a so-called {\it explicit valence-bond crystal
state},\cite{lhuillier03} which respects the lattice symmetry.

Although we could expect a gapped quantum paramagnetic {
explicit valence-bond crystal} GS for the
star lattice in case of strong dimer bonds $J_D \gg J_T$,
this
should be contrasted with other models where an {
explicit valence-bond crystal} GS arises in the limit of strong dimer bonds.
For example, in the simple $s=1/2$ Heisenberg bilayer model, the picture of
weakly coupled dimers is qualitatively correct only for an interlayer
exchange coupling $J_\perp$ significantly larger than the
intralayer coupling $J$ ($J_\perp \gtrsim 2.5 J$, see
Refs.\ \onlinecite{gwr95,bilayer} and references therein).
Because the bilayer Heisenberg model is not frustrated,
the classical GS does not exhibit any non-trivial degeneracy
and the GS remains the semi-classical \Neel state for
$J_\perp \approx J$ and all values of $s$.\cite{bilayer}
By contrast, the quantum paramagnetic GS appears in the
$s=1/2$ star lattice even in the {\it uniform} case $J_D=J_T$.
This difference can be attributed to the strong frustration
present in the star lattice.

The magnetization curve of the $s=1/2$ HAFM on the star lattice
shows a jump just below saturation
and several candidates for plateaux e.g.\ at $m=1/3$, $7/9$ and $8/9$ 
as some typical features of strongly frustrated quantum spin lattices. 
Furthermore, low-energy excitations present for certain magnetic fields
promise large magnetocaloric effects.\cite{entropy}

\begin{acknowledgments}

We are indebted to H.-U.~Everts for valuable discussions.
This work was partly supported by the DFG (project Ri615/12-1).

\end{acknowledgments}



\begin{thebibliography}{99}

\bibitem{taniguchi95}
  S.~Taniguchi\etal{, T.~Nishikawa, Y.~Yasui, Y.~Kobayashi, M.~Sato,
  T.~Nishioka, M.~Kontani, K.~Sano}, 
  {J. Phys. Soc. Jpn.} \textbf{64}, 2758 (1995).

\bibitem{kageyama99}
 H.~Kageyama\etal{, K.~Yoshimura, R.~Stern, N.V.~Mushnikov, K.~Onizuka,
 M.~Kato, K.~Kosuge, C.P.~Slichter, T.~Goto, Y.~Ueda},
 {Phys. Rev. Lett.} \textbf{82}, 3168 (1999).

\bibitem{coldea}
R.~Coldea, D.A.~Tennant, Z.~Tylczynski,
Phys. Rev. B {\bf 68,} 134424 (2003).

\bibitem{TOKMIG02} H.~Tanaka\etal{, T.~Ono, H.A.~Katori, H.~Mitamura,
   F.~Ishikawa, T.~Goto}, Progr. Theor. Phys. Suppl. {\bf 145}, 101 (2002).

\bibitem{gruenbaum}
 B.~Gr\"unbaum, G.C.~Shephard, \textit{Tilings and Patterns},
 W.H.~Freeman and Company, New York (1987). 

\bibitem{phd} J.~Schulenburg,
  PhD thesis, University of Magdeburg (2002) \par
  [\verb"http://www-e.uni-magdeburg.de/jschulen/diss.html"].

\bibitem{wir04}
J.~Richter, J.~Schulenburg, A.~Honecker, 
  {Lect.\ Notes Phys.\ {\bf 645}, 85-153 (2004)}.

\bibitem{lhuillier03}
G.~Misguich,   C.~Lhuillier,
{\it Two-dimensional quantum antiferromagnets},
 \verb"cond-mat/0310405" (2003).

\bibitem{lhuillier01}
C.~Lhuillier, P.~Sindzingre, J.-B.~Fouet,
Can. J. Phys. {\bf 79}, 1525 (2001).

\bibitem{moessner01}
R.~Moessner,
Can. J. Phys. {\bf 79}, 1283 (2001).

\bibitem{schmalfuss02}
 D.~Schmalfu{\ss}\etal{, P.~Tomczak, J.~Schulenburg, J.~Richter},
 {Phys. Rev. B} \textbf{65}, 224405 (2002).

\bibitem{tomczak99}
 P.~Tomczak, J.~Richter,
 {Phys. Rev. B} \textbf{59}, 107 (1999).

\bibitem{tomczak01}
 P.~Tomczak\etal{, J.~Schulenburg, J.~Richter, A.R.~Ferchmin},
 {J. Phys.: Condens. Matter} \textbf{13}, 3851 (2001).

\bibitem{normand97}
 B.~Normand\etal{, K.~Penc, M.~Albrecht, F.~Mila},
 {Phys. Rev. B} \textbf{56}, R5736 (1997).

\bibitem{miyahara98}
 S.~Miyahara\etal{, M.~Troyer, D.C.~Johnston, K.~Ueda},
 {J. Phys. Soc. Jpn.} \textbf{67}, 3918 (1998).

\bibitem{harris92}
 A.B.~Harris, C.~Kallin, A.J.~Berlinsky,
 {Phys. Rev. B} \textbf{45}, 2899 (1992).

\bibitem{ChaHoShe}
 J.T. Chalker, P C.W. Holdsworth, E.F. Shender,
 Phys. Rev. Lett. {\bf 68}, 855 (1992); 
 E.F. Shender, P.C.W. Holdsworth,
 In {\it Fluctuations and Order: a new synthesis},
 M.M. Millonas ed., Springer-Verlag (1996).

\bibitem{asakawa94} 
H. Asakawa, M. Suzuki, Physica A {\bf 205}, 687 (1994).

\bibitem{betts98}
 D.D.~Betts\etal{, J.~Schulenburg, G.E.~Stewart, J.~Richter, J.S.~Flynn},
 {J. Phys. A: Math. Gen.} \textbf{31}, 7685 (1998).

\bibitem{leung93}
 P.W.~Leung, V.~Elser,
 {Phys. Rev. B} \textbf{47}, 5459 (1993).

\bibitem{bernu94}
 B.~Bernu\etal{, P.~Lecheminant, C.~Lhuillier, L.~Pierre}, 
 {Phys. Rev. B} \textbf{50}, 10048 (1994).

\bibitem{lecheminant97}
 P.~Lecheminant\etal{, B.~Bernu, C.~Lhuillier, L.~Pierre, P.~Sindzingre},
 {Phys. Rev. B} \textbf{56}, 2521 (1997).

\bibitem{waldtmann98}
 Ch.~Waldtmann\etal{, H.-U.~Everts, B.~Bernu, C.~Lhuillier, P.~Sindzingre,
 P.~Lecheminant, L.~Pierre}, {Eur. Phys. J. B} \textbf{2}, 501 (1998).

\bibitem{neuberger89}
 H.~Neuberger, T.~Ziman, {Phys. Rev. B} \textbf{39}, 2608 (1989).

\bibitem{hasenfratz93}
 P.~Hasenfratz, F.~Niedermayer,
 {Z. Phys. B} \textbf{92}, 91 (1993).

\bibitem{gwr95} C. Gros, W. Wenzel, J. Richter, Europhys. Lett. {\bf 32},
 747 (1995).

\bibitem{krsfb00} S.E.~Kr\"uger\etal{, J.~Richter, J.~Schulenburg,
 D.J.J.~Farnell, R.F.~Bishop}, Phys. Rev. B {\bf 61}, 14607 (2000). 

\bibitem{kagplateau} D.C.~Cabra\etal{, M.D.~Grynberg, P.C.W.~Holdsworth,
   A.~Honecker, P.~Pujol, J.~Richter, D.~Schmalfu{\ss}, J.~Schulenburg},
             \verb"cond-mat/0404279" (2004).

\bibitem{jump} J. Schulenburg\etal{, A. Honecker, J. Schnack, J. Richter,
                H.-J. Schmidt}, Phys. Rev. Lett. {\bf 88}, 167207 (2002);
                J. Richter\etal{, J. Schulenburg, A. Honecker, J. Schnack,
                H.-J. Schmidt}, 
                J. Phys.: Condens. Matter {\bf 16}, S779 (2004). 

\bibitem{entropy} O. Derzhko, J. Richter, Phys. Rev. B {\bf 70}, 104415 (2004);
                M.E. Zhitomirsky, A. Honecker, J. Stat. Mech.: Theor. Exp.
                P07012 (2004);
                M.E. Zhitomirsky, H. Tsunetsugu,
                Phys. Rev. B {\bf 70}, 100403(R) (2004).

\bibitem{sp04} J. Richter, O. Derzhko, J. Schulenburg,
               Phys. Rev. Lett. {\bf 93}, 107206 (2004).

\bibitem{bilayer} M.P. Gelfand\etal{, Zheng Weihong, C.J. Hamer, J. Oitmaa},
                Phys. Rev. B {\bf 57}, 392 (1998);
                M. Troyer, S. Sachdev, Phys. Rev. Lett. {\bf 81}, 5418 (1998).


\end{thebibliography}
\end{document}